\documentclass[aps,preprint]{revtex4}%
\usepackage{amsfonts}
\usepackage{amsmath}
\usepackage{amssymb}
\usepackage{graphicx}
\usepackage{array}
\usepackage{graphicx}
\usepackage{booktabs}
\usepackage{multirow}
\usepackage{makecell}
\usepackage{float}%
\begin{document}
\title{Chaos bound and its violation in charged Kiselev black hole}

\author{Chuanhong Gao}
\email{chuanhonggao@hotmail.com}
\author{Deyou Chen}
\email{deyouchen@hotmail.com}
\affiliation{School of Science, Xihua University, Chengdu 610039, China}

\author{Peng Wang}
\email{pengw@scu.edu.cn}
\affiliation{Center for Theoretical Physics, College of Physics, Sichuan University, Chengdu, 610064, China}

\author{Chengye Yu}
\email{chengyeyu1@hotmail.com}
\affiliation{School of Science, Xihua University, Chengdu 610039, China}

\begin{abstract}
{The chaos bound in the near-horizon regions has been studied through the expansions of the metric functions on the horizon. In this paper, we investigate the chaos bound in the the near-horizon region and at a certain distance from the horizon of a charged Kiselev black hole. The value of the Lyapunov exponent is accurately calculated by a Jacobian matrix. The angular momentum of a charged particle around the black hole affects not only the exponent, but also the position of the equilibrium orbit. This position gradually moves away from the horizon with the increase of the angular momentum. We find that the the bound is violated at a certain distance from the horizon and there is no violation in the near-horizon region when the charge mass ratio of the particle is fixed. The small value of the normalization factor is more likely to cause the violation.}
\end{abstract}
\maketitle
\tableofcontents

\section{Introduction}

Chaos is a kind of unpredictable and random motion in deterministically and nonlinearly dynamic systems due to the sensitivity to initial conditions. In quantum many-body systems, an important probe of chaotic behaviors is out-of-time ordered correlators (OTOCs). Through the diagnosis of the chaos, people found that it increases exponentially with time, $C(t)\approx exp(\lambda t)$, where $\lambda$ is a Lyapunov exponent to characterize the sensitive dependence on the initial conditions \cite{HS1,HS2,RRP,JKY,LR}.

Recently, Maldacena, Shenker and Stanford researched the chaos in the thermal quantum system with a large number of degrees of freedom by using an OTOC. Based on the factorization assumption and the assumption that there is a large hierarchy between the dissipation time and the scrambling time, they put forward a conjecture that there is a universal upper bound for the Lyapunov exponent,

\begin{eqnarray}
\lambda \leq \frac{2\pi T}{\hbar},
\label{eq1.1}
\end{eqnarray}

\noindent where $T$ is the temperature of the system \cite{MSS}. After this seminal conjecture was put forward, it immediately attracted much attention. The chaos in various models and gravitational theories was researched \cite{PR,MS,ADNT,GHS,GZZ,AKY,JASA,GZ,CFCZZ,HT,KS,BINT,HG,LTW1,CMBWZ,MWZ,PP1,PP2,CWJ1,WCJ2,MWZ1,LTW2,LW,CKR,DMM1,DMM2,CKR1,CKR2,ACO,MPD,HY,GR,CCKM1,BL,MRS,QCL,DG}. These researches play an important role in black hole physics and quantum information. When a particle is subjected to sufficiently strong electromagnetic or scalar forces, the particle can be very close to a black hole without falling into it. At this time, the chaotic behavior of the particle was discussed in \cite{HT}. The value of the Lyapunov exponent was gotten and equal to the surface gravity of the black hole. From the relation between the surface gravity and temperature, this result supports the conjecture of Maldacena et al.

In the recent work, some violations of the chaos bound were found \cite{ZLL,KG1,KG2,LG1,LG2}. The static equilibrium of a charged probe particle around a black hole can be provided by the Lorentz force. Adjusting the charge mass ratio of the particle make it infinitely close to the event horizon. Taking into account the contributions of the sub-leading terms in the near-horizon expansion, Zhao et al studied the chaos bound in the near-horizon regions by using the effective potentials \cite{ZLL}. They found that the bound was satisfied by the Reissner-Nordstr$\ddot{o}$m and Reissner-Nordstr$\ddot{o}$m-AdS black holes and violated by a large number of charged black holes. In the derivation, they only considered the contribution of the radial directions. In fact, the angular momentum of the particle has an influence on the Lyapunov exponent. This is because the angular momentum affects the effective potential and increases the magnitude of the chaotic behavior of the particle. When this influence is considered, the chaos bound in the near-horizon regions of the Reissner-Nordstr$\ddot{o}$m and Reissner-Nordstr$\ddot{o}$m-AdS black holes was studied again \cite{LG2}. It was found that the bound is violated in the near-horizon regions when the charge of the black holes and the angular momentum of the particle are large. In the rotating charged black holes, the violation of the bound was also found through the calculation of the effective potentials \cite{KG1,KG2}.

In this paper, we investigate the influence of the angular momentum of a charged particle on the chaos bound through circular motions of the particle around a charged Kiselev black hole. This black hole describes a space-time surrounded by an anisotropic fluid. It was first gotten by Kiselev \cite{BL1} and its properties were studied in \cite{MV1,MV2}. Based on the work of Kiselev, some exact black hole solutions were gotten in \cite{BL3,BL4,TD2} and their properties were discussed in \cite{OFH,YJBH,CFO,KO1,KO2,BL6,GM1,GM2}. Here the value of the Lyapunov exponent is accurately derived by a Jacobian matrix. A position of a equilibrium orbit of the particle is affected by its charge mass ratio and angular momentum. By fixing the charge mass ratio and changing the value of the angular momentum, we obtain positions of different equilibrium orbits. The bound is numerically discussed in the near-horizon region and at a certain distance from the horizon.

The rest is organized as follows. In the next section, taking into account a motion of a charged particle in the equatorial plane of a spherically symmetric black hole, we obtain a general expression of Lyapunov exponent by calculating the eigenvalue of the Jacobian matrix. In Section 3, we investigate the influence of the angular momentum of the particle on the chaos bound in the near-horizon region and at a certain distance from the horizon of the charged Kiselev black hole. Section 4 is devoted to our conclusions.

\section{Lyapunov exponent in charged black holes}

We consider a circular motion of a charged particle in the equatorial plane of a spherically symmetric black hole to obtain the Lyapunov exponent. The black hole is given by

\begin{eqnarray}
ds^2 = -F(r)dt^2 + \frac{1}{N(r)}dr^2 + C(r)d\theta^2+D(r)d\phi^2,
\label{eq2.1}
\end{eqnarray}

\noindent with an electromagnetic potential $A_{\mu}=A_tdt$. When the particle with charge $q$ moves around the black hole, its Lagrangian is

\begin{eqnarray}
\mathcal{L} = \frac{1}{2}\left(-F\dot{t}^2+\frac{\dot{r}^2}{N} +D\dot{\phi}^2\right) -qA_t\dot{t},
\label{eq2.2}
\end{eqnarray}

\noindent where $\dot{x^{\mu}} = \frac{dx^{\mu}}{d\tau}$ and $\tau$ is a proper time. Using the definition of the generalized momentum $ \pi_{\mu}=\frac{\partial\mathcal{L}}{\partial\dot{x}}$, we get

\begin{eqnarray}
\pi_t = -F\dot{t} -qA_t=-E, \quad\quad \pi_r = \frac{\dot{r}}{N}, \quad\quad \pi_{\phi} = D \dot{\phi}=L.
\label{eq2.3}
\end{eqnarray}

\noindent In the above equation, $E$ and $L$ denote the energy and angular momentum of the particle, respectively. Thus the Hamiltonian of the particle is

\begin{eqnarray}
H = \frac{-(\pi_{t}+qA_{t})^2+\pi_r^2FN+ \pi^2_{\phi}D^{-1}F}{2F}.
\label{eq2.4}
\end{eqnarray}

\noindent From the Hamiltonian, the equations of motion are gotten, which are

\begin{eqnarray}
\dot{t} &=& \frac{\partial H}{\partial \pi_t}=-\frac{\pi_t+qA_t}{F}, \quad  \dot{\pi_t}= -\frac{\partial H}{\partial t} =0 ,
\quad \dot{r} = \frac{\partial H}{\partial \pi_r}= \pi_r N, \nonumber\\
\dot{\pi_r} &=& -\frac{\partial H}{\partial r} =-\frac{1}{2}\left[\pi^2_r N^{\prime} -\frac{2qA^{\prime}_t(\pi_t+qA_t)}{F}+\frac{(\pi_t+qA_t)^2F^{\prime}}{F^2} -\pi^2_{\phi}D^{-2}D^{\prime}\right], \nonumber\\
\dot{\phi} &=& \frac{\partial H}{\partial \pi_{\phi}}= \frac{\pi_{\phi}}{D}, \quad  \dot{\pi_\phi}= -\frac{\partial H}{\partial \phi} =0.
\label{eq2.5}
\end{eqnarray}

\noindent In the above equations, "$\prime$" represents the derivative of $r$. Using the equations of motion, we get the relations between the radial coordinate and time and between the radial momentum and time

\begin{eqnarray}
\frac{dr}{dt} &=& \frac{\dot{r}}{\dot{t}} =-\frac{\pi_rFN}{\pi_t+qA_t}, \nonumber\\
\frac{d\pi_r}{dt} &=& \frac{\dot{\pi_r}}{\dot{t}} = -qA^{\prime}_t +\frac{1}{2}\left[\frac{\pi^2_r FN^{\prime}}{\pi_t+qA_t}+\frac{(\pi_t+qA_t)F^{\prime}}{F} -\frac{\pi^2_{\phi}D^{-2}D^{\prime}F}{\pi_t+qA_t}\right].
\label{eq2.6}
\end{eqnarray}

\noindent For convenience, we define $F_1= \frac{dr}{dt}$ and $F_2=\frac{d\pi_r}{dt}$. The normalization of the four-velocity of a particle is given by $ g_{\mu\nu}\dot{x}^{\mu}\dot{x}^{\nu}=\eta$, where $\eta =0$ describes the case of a photon, and $\eta =-1$ corresponds to the case of a massive particle. Here the particle is charged. Using the normalization and the metric (\ref{eq2.1}) yields a constrain condition

\begin{eqnarray}
\pi_t+qA_t=-\sqrt{F(1+ \pi_r^2N + \pi_{\phi}^2D^{-1})}.
\label{eq2.7}
\end{eqnarray}

\noindent We use this constrain and rewrite Eq. (\ref{eq2.6}) as

\begin{eqnarray}
F_1 &=& \frac{\pi_rFN}{\sqrt{F(1+ \pi_r^2N + \pi_{\phi}^2D^{-1})}}, \nonumber\\
F_2 &=& -qA^{\prime}_t -\frac{\pi^2_r (NF)^{\prime}+F^{\prime}}{2\sqrt{F(1+ \pi_r^2N + \pi_{\phi}^2D^{-1})}} -\frac{\pi^2_{\phi}(D^{-1}F)^{\prime}}{2\sqrt{F(1+ \pi_r^2N + \pi_{\phi}^2D^{-1})}}.
\label{eq2.8}
\end{eqnarray}

The effective potential of the particle plays an important role in the acquisition of the Lyapunov exponent \cite{ZLL,KG1,KG2}. Here the Lyapunov exponent is derived by the eigenvalue of a Jacobian matrix in the phase space $(r, \pi_r)$. In this phase space, the Jacobian matrix is defined by $K_{ij}$ and the matrix elements are

\begin{eqnarray}
K_{11} &=&\frac{\partial F_1}{\partial r} = \frac{\pi_r(NF)^{\prime}}{\sqrt{F(1+ \pi_r^2N + \pi_{\phi}^2D^{-1})}}  -\pi_rNF \frac{F^{\prime}+\pi_r^2(NF)^{\prime}+\pi_{\phi}^2(D^{-1}F)^{\prime}}{2\left[F(1+ \pi_r^2N + \pi_{\phi}^2D^{-1})\right]^{3/2}}, \nonumber\\
K_{12}&=& \frac{\partial F_1}{\partial \pi_r} =\frac{FN}{\sqrt{F(1+ \pi_r^2N + \pi_{\phi}^2D^{-1})}}-\frac{\pi_r^2F^2N^2}{\left[F(1+ \pi_r^2N + \pi_{\phi}^2D^{-1})\right]^{3/2}},\nonumber\\
K_{21}&=& \frac{\partial F_2}{\partial \pi_r} = -qA^{\prime\prime}_t
-\frac{F^{\prime\prime}+\pi^2_r(NF)^{\prime\prime}+ \pi^2_{\phi}(D^{-1}F)^{\prime\prime}}{2\sqrt{F(1+ \pi_r^2N + \pi_{\phi}^2D^{-1})}} + \frac{\left[F^{\prime}+\pi^2_r(NF)^{\prime} +\pi_{\phi}^2(D^{-1}F)^{\prime}\right]^2}{4\left[F(1+ \pi_r^2N + \pi_{\phi}^2D^{-1})\right]^{3/2}},\nonumber\\
K_{22} &=& \frac{\partial F_2}{\partial \pi_r} = -\frac{\pi_r(NF)^{\prime}}{\sqrt{F(1+ \pi_r^2N + \pi_{\phi}^2D^{-1})}}+ \frac{\pi_rNF}{2}\frac{F^{\prime}+\pi^2_r(NF)^{\prime}+\pi^2_{\phi}(D^{-1}F)^{\prime}}{\left[F(1+ \pi_r^2N + \pi_{\phi}^2D^{-1})\right]^{3/2}}.
\label{eq2.9}
\end{eqnarray}

\noindent Taking into account the motion of the particle in a equilibrium orbit, we use a condition $\pi_r=\frac{d\pi_r}{dt}=0$ to constrain the trajectory of the particle. Using the constrain and calculating the eigenvalue, we get the exponent

\begin{eqnarray}
\lambda^2 = \frac{1}{4}\frac{N\left[F^{\prime}+\pi_{\phi}^2(D^{-1}F)^{\prime}\right]^2}{F(1+\pi_{\phi}^2D^{-1})^2} -\frac{1}{2}N\frac{F^{\prime\prime}+\pi_{\phi}^2(D^{-1}F)^{\prime\prime}}{1+\pi_{\phi}^2D^{-1}}
-\frac{qA_t^{\prime\prime}FN}{\sqrt{F(1+\pi_{\phi}^2D^{-1})}}.
\label{eq2.10}
\end{eqnarray}

\noindent It is clearly that the exponent is affected by the angular momentum. When $\pi_{\phi }=0$, it implies that the contribution of the angular momentum is neglected. Since the contribution of the angular momentum plays an important role in the Lyapunov exponent, we do no neglect it in this paper.

\section{Chaos bound and its violation in Kiselev black hole}

We use a probe particle with mass $m$ moving around a charged Kiselev black hole to investigate the chaos bound. When the particle is neutral, a force on it is a centrifugal force, and this force is related to the angular momentum of the particle. This angular momentum causes the change of the effective potential, but the angular momentum is not sufficient to make the particle's orbit closer to the horizon \cite{HT}. When the particle is charged with $q$, one can adjust the charge mass ratio to make it close to or away from the horizon. In \cite{ZLL,LG2}, the authors studied the violation of the bound in the near-horizon regions through the expansions at the event horizon. Here we investigate the influence of the angular momentum on the bound in the near-horizon region and at a certain distance from the horizon.

The metric of the black hole is given by the metric (\ref{eq2.1}), where

\begin{eqnarray}
F(r)=N(r)=1-\frac{2M}{r}+\frac{Q^{2}}{r^{2}}-\frac{\alpha}{r^{3\omega +1}}, \quad\quad C(r)=r^{2}, \quad\quad D(r)=r^{2}sin^2\theta,
\label{eq3.1}
\end{eqnarray}

\noindent $M$ and $Q$ are the mass and electric charge of the black hole, respectively. $\alpha$ is a normalization factor and $\omega$ is a state parameter characterizing the anisotropic fluid. The metric (\ref{eq2.1}) describes an asymptotically flat solution when $-\frac{1}{3}\le\omega<0$ and a non-asymptotically flat solution when $-1<\omega<-\frac{1}{3}$. Some classical black hole solutions can be obtained by the specific values of $\omega$ and $\alpha$. The metric of Reissner-Nordstr$\ddot{o}$m-(Anti)-de Sitter black holes is recovered when $\omega = -1$ and $3 \alpha$ plays the role of the cosmological constant. When $\omega = -\frac{1}{3}$, the metric describes a topological Reissner-Nordstr$\ddot{o}$m solution. When $\alpha = 0$, it describes a  Reissner-Nordstr$\ddot{o}$m solution. Although $\omega$ can have many values, our interest is focused on the special cases where $\omega=-\frac{1}{2}$ and $\omega=-\frac{2}{3}$.  Its electromagnetic potential is $A_t=\frac{Q}{r}$. The event horizon ($r_+$) is determined by $F(r)= 0 $. In the equatorial plane, we get $\theta = \frac{\pi}{2}$ and $D(r)=r^{2}$.

\begin{table}[H]
\begin{center}
\setlength{\tabcolsep}{2.5mm}
\begin{tabular}{ccccccccc}
\toprule[1pt]
&  L &  0 &  1 &  2 & 3 &  5 &  10 &  20 \\  \Xcline{2-9}{0.3pt}
\multirow{3}*{$r_0$} & $\alpha$=0.020  & 1.378386 & 1.379215 & 1.381689 & 1.385770 & 1.398472 &  1.451301 & 1.588752  \\   \Xcline{2-9}{0.3pt}
&$\alpha$=0.060  & 1.519514 & 1.520442 & 1.523206 & 1.527749 & 1.541780 & 1.598667 & 1.741000  \\   \Xcline{2-9}{0.3pt}
&$\alpha$=0.096  & 1.666155 & 1.667152 & 1.670118 & 1.674985 & 1.689946  & 1.749770 & 1.749770  \\
\bottomrule[1pt]
\end{tabular}
\label{tab1}
\end{center}
Table 1. The position of the equilibrium orbit changes with the value of the angular momentum of the particle when $\omega=-\frac{1}{2}$ and $Q=0.95$. The event horizon is located at $r_+ =1.376809$ when $\alpha=0.020$, at $r_+=1.517366$ when $\alpha=0.060$ and at $r_+=1.663377$ when $\alpha=0.096$.
\end{table}

\begin{table}[H]
\begin{center}
\setlength{\tabcolsep}{2.5mm}
\begin{tabular}{ccccccccc}
\toprule[1pt]
& L & 0 & 1 & 2 & 3 & 5 & 10 & 20 \\  \Xcline{2-9}{0.4pt}
\multirow{3}*{$r_0$} & $\alpha$=0.060 & 1.399487 & 1.400110 & 1.401974 & 1.405067 & 1.414827 & 1.457618  & 1.584125  \\   \Xcline{2-9}{0.4pt}
&$\alpha$=0.090 & 1.531051 & 1.531177 & 1.533915 & 1.543746 & 1.548549 & 1.595668  & 1.726790  \\   \Xcline{2-9}{0.4pt}
&$\alpha$=0.120& 1.677243 & 1.678035 & 1.680399 & 1.684295 & 1.696411  & 1.746872 & 1.882388 \\
\bottomrule[1pt]
\end{tabular}
\label{tab1}
\end{center}
Table 2. The position of the equilibrium orbit changes with the value of the angular momentum of the particle when $\omega=-\frac{1}{2}$ and $Q=0.99$. The event horizon is located at $r_+=1.398266$ when $\alpha=0.060$, at $r_+=1.529366$ when $\alpha=0.090$ and at $r_+=1.675008$ when $\alpha=0.120$.
\end{table}

\begin{table}[H]
\begin{center}
\setlength{\tabcolsep}{2.5mm}
\begin{tabular}{ccccccccc}
\toprule[1pt]
& L & 0 & 1 & 2 & 3 & 5 & 10 & 20 \\  \Xcline{2-9}{0.4pt}
\multirow{3}*{$r_0$} & $\alpha$=0.020 & 1.390238 & 1.391053 & 1.393485 & 1.397495 & 1.409974 & 1.461827  & 1.596804  \\   \Xcline{2-9}{0.4pt}
&$\alpha$=0.060 & 1.579006 & 1.579873 & 1.582457 & 1.586702 & 1.599804 & 1.652856  & 1.786028  \\   \Xcline{2-9}{0.4pt}
&$\alpha$=0.096 & 1.828424 & 1.829270 & 1.831788 & 1.835921 & 1.848626  & 1.899553  & 2.026144 \\
\bottomrule[1pt]
\end{tabular}
\label{tab1}
\end{center}
Table 3. The position of the equilibrium orbit changes with the value of the angular momentum of the particle when $\omega=-\frac{2}{3}$ and $Q=0.95$. The event horizon is located at $r_+=1.388661$ when $\alpha=0.020$, at $r_+=1.576836$ when $\alpha=0.060$ and at $r_+=1.825580$ when $\alpha=0.096$.
\end{table}

\begin{table}[H]
\begin{center}
\setlength{\tabcolsep}{2.5mm}
\begin{tabular}{ccccccccc}
\toprule[1pt]
& L & 0 & 1 & 2 & 3 & 5 & 10 & 20 \\  \Xcline{2-9}{0.4pt}
\multirow{3}*{$r_0$} & $\alpha$=0.060 & 1.451859 & 1.452452 & 1.454266 & 1.457163 & 1.461238 &  1.506609 & 1.624338 \\   \Xcline{2-9}{0.4pt}
&$\alpha$=0.090  & 1.654750 & 1.655394 & 1.657314 & 1.660482 & 1.670363 & 1.711977 & 1.826890  \\   \Xcline{2-9}{0.4pt}
&$\alpha$=0.120 & 1.965787 & 1.966395 & 1.968206 & 1.971190 & 1.980453  & 2.018876 & 2.122473 \\
\bottomrule[1pt]
\end{tabular}
\label{tab1}
\end{center}
Table 4. The position of the equilibrium orbit changes with the value of the angular momentum of the particle when $\omega=-\frac{2}{3}$ and $Q=0.99$. The event horizon is located at $r_+=1.450608$ when $\alpha=0.060$, at $r_+=1.652985$ when $\alpha=0.090$ and at $r_+=1.963428$ when $\alpha=0.120$.
\end{table}

We first find the positions of the equilibrium orbits. When $\omega =-\frac{1}{2}$, Eq. (\ref{eq3.1}) takes form

\begin{eqnarray}
F(r)=1-\frac{2M}{r}+\frac{Q^{2}}{r^{2}}-\alpha\sqrt{r},
\label{eq3.2}
\end{eqnarray}

\noindent and then the surface gravity is $\kappa=\frac{M}{r_+^2}-\frac{Q^2}{r_+^3}-\frac{\alpha}{4\sqrt{r_+}}$. When $\omega=-\frac{2}{3}$, Eq. (\ref{eq3.1}) becomes

\begin{eqnarray}
F(r)=1-\frac{2M}{r}+\frac{Q^{2}}{r^{2}}-\alpha r,
\label{eq3.3}
\end{eqnarray}

\noindent and the surface gravity is $\kappa=\frac{M}{r_+^2}-\frac{Q^2}{r_+^3}-\frac{\alpha }{2}$. Using the condition $\pi_r=\frac{d\pi_r}{dt}=0$ and Eqs. (\ref{eq2.8}), (\ref{eq3.2}) and (\ref{eq3.3}), we get the expressions of the equilibrium orbits. Due to their complexity, the specific positions $r_0$ of the orbits are numerically analyzed. We order $M = 1$, $m = 1$, $q = 15$ in this paper, and get the positions of some certain orbits in Table 1-Table 4. The location of the horizon is also obtained by numerical solutions. From the tables, it is clearly found that when the state parameter, electric charge and normalization factor are fixed, the positions of the orbits gradually move away from the horizon with the increase of the angular momentum. The equilibrium orbits are very close to the horizon when the angular momentum is small ($L < 5$). When the state parameter, electric charge and angular momentum are fixed, the values of the positions of the orbits and horizon increase with the increase of the value of the normalization factor.

\begin{figure}[h]
\centering
\includegraphics[width=12cm,height=9cm]{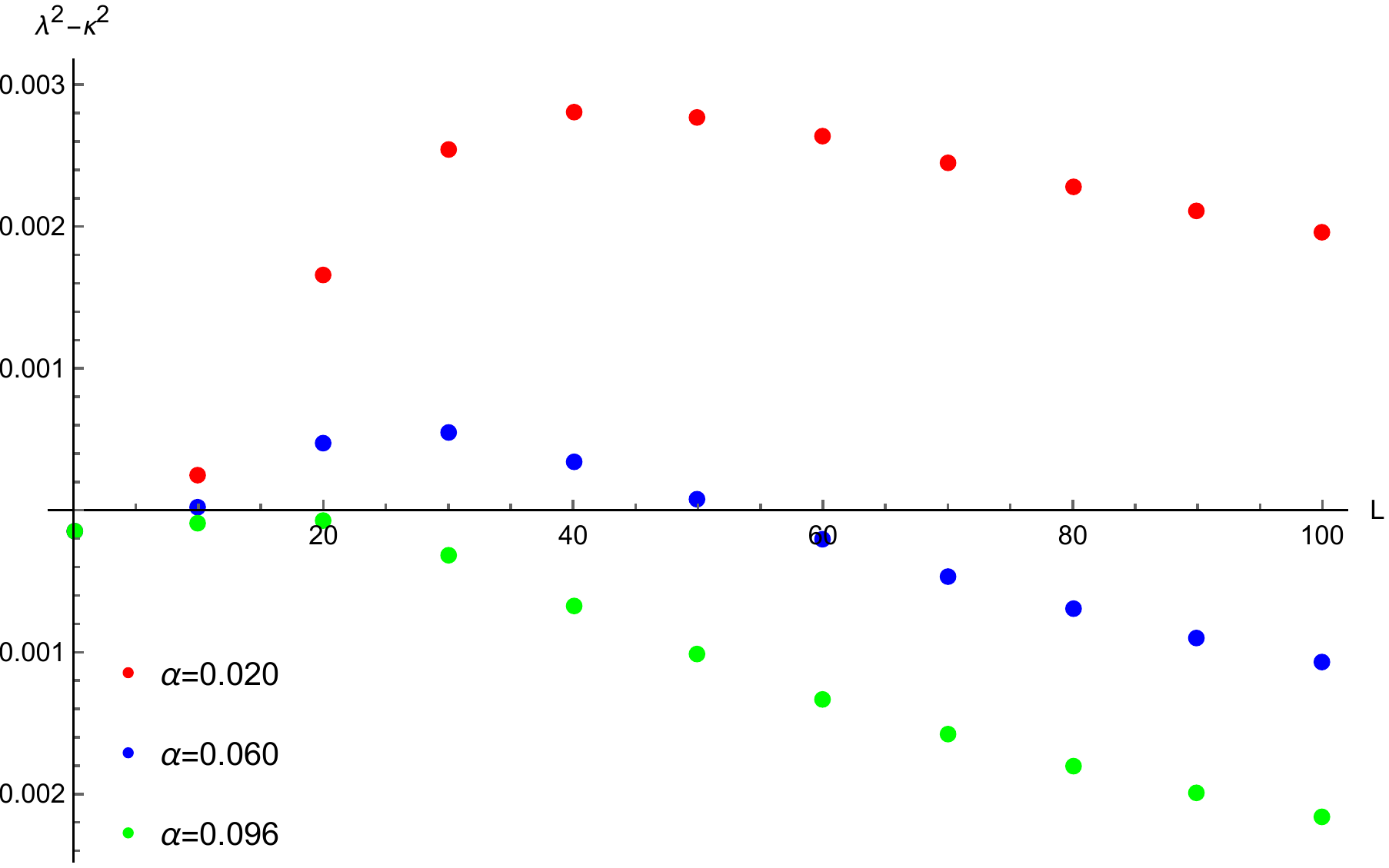}
\caption{The influence of the angular momentum on the chaos bound, where $\omega=-\frac{1}{2}$ and $Q=0.95$. The violation of the bound occurs at $L>7.25$ when $\alpha=0.020$ and at $9.27<L<52.41$ when $\alpha=0.060$. When $\alpha=0.096$, there is no violation.}
\end{figure}

\begin{figure}[h]
\centering
\includegraphics[width=12cm,height=9cm]{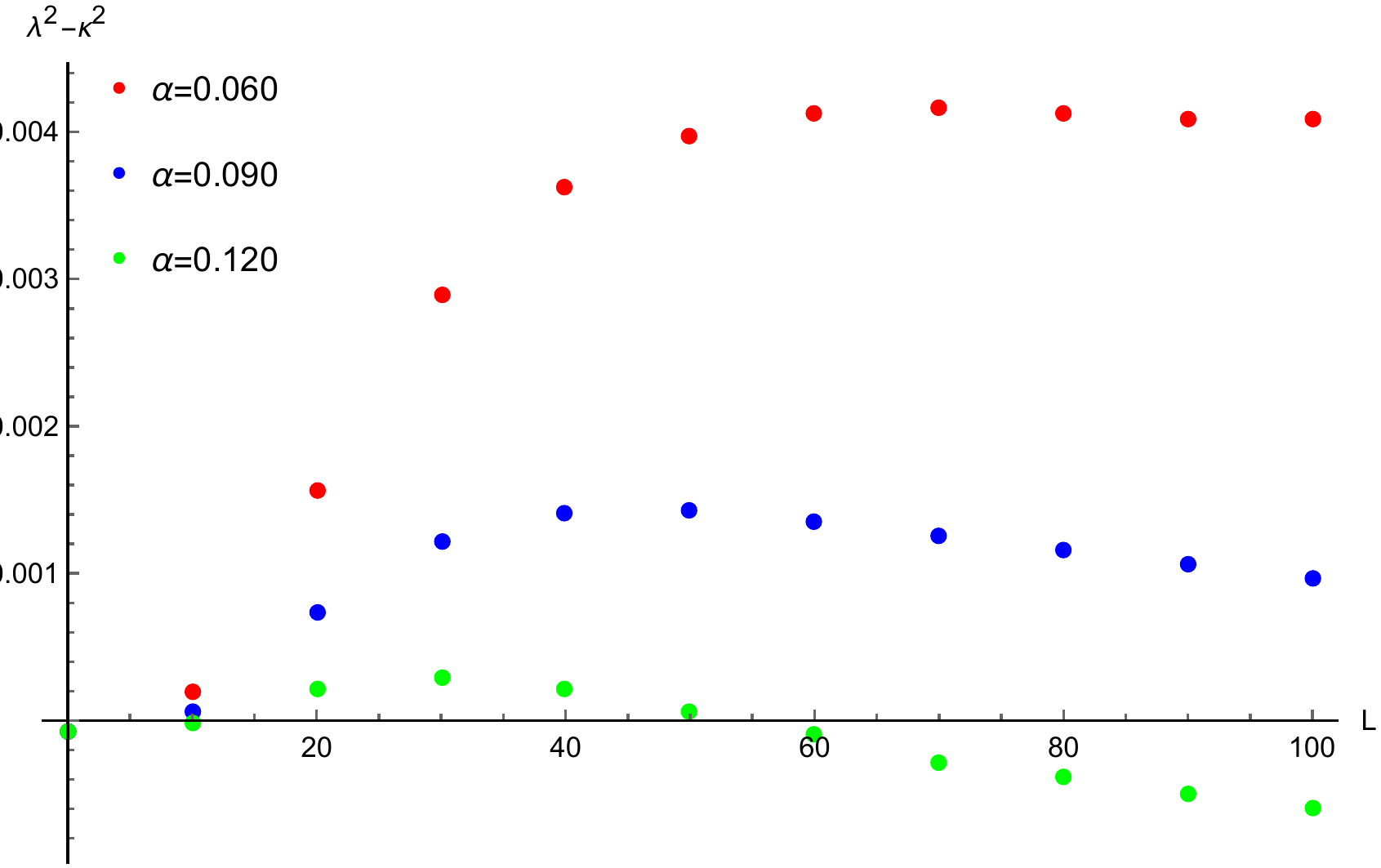}
\caption{The influence of the angular momentum on the chaos bound, where $\omega=-\frac{1}{2}$ and $Q=0.99$. The violation of the bound occurs at $L>6.85$ when $\alpha=0.060$, at $L>8.25$ when $\alpha=0.090$ and at $10.51<L<58.95$ when $\alpha=0.120$.}
\end{figure}

\begin{figure}[h]
\centering
\includegraphics[width=12cm,height=9cm]{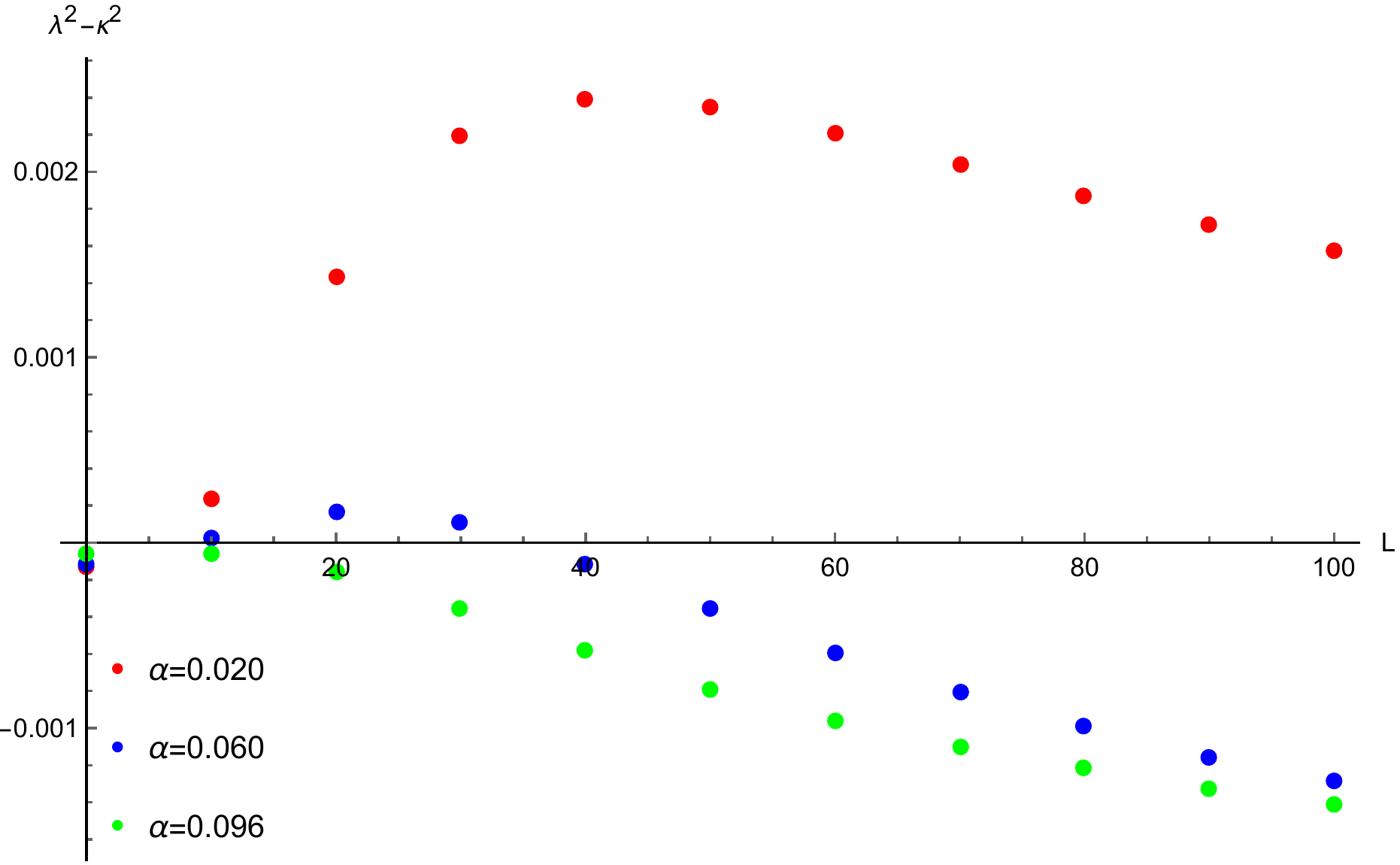}
\caption{The influence of the angular momentum on the chaos bound, where $\omega=-\frac{2}{3}$ and $Q=0.95$. The violation of the bound occurs at $L>7.42$ when $\alpha=0.020$ and at $10.91<L<35.28$ when $\alpha=0.060$. When $\alpha=0.096$, there is no violation.}
\end{figure}

\begin{figure}[h]
\centering
\includegraphics[width=12cm,height=9cm]{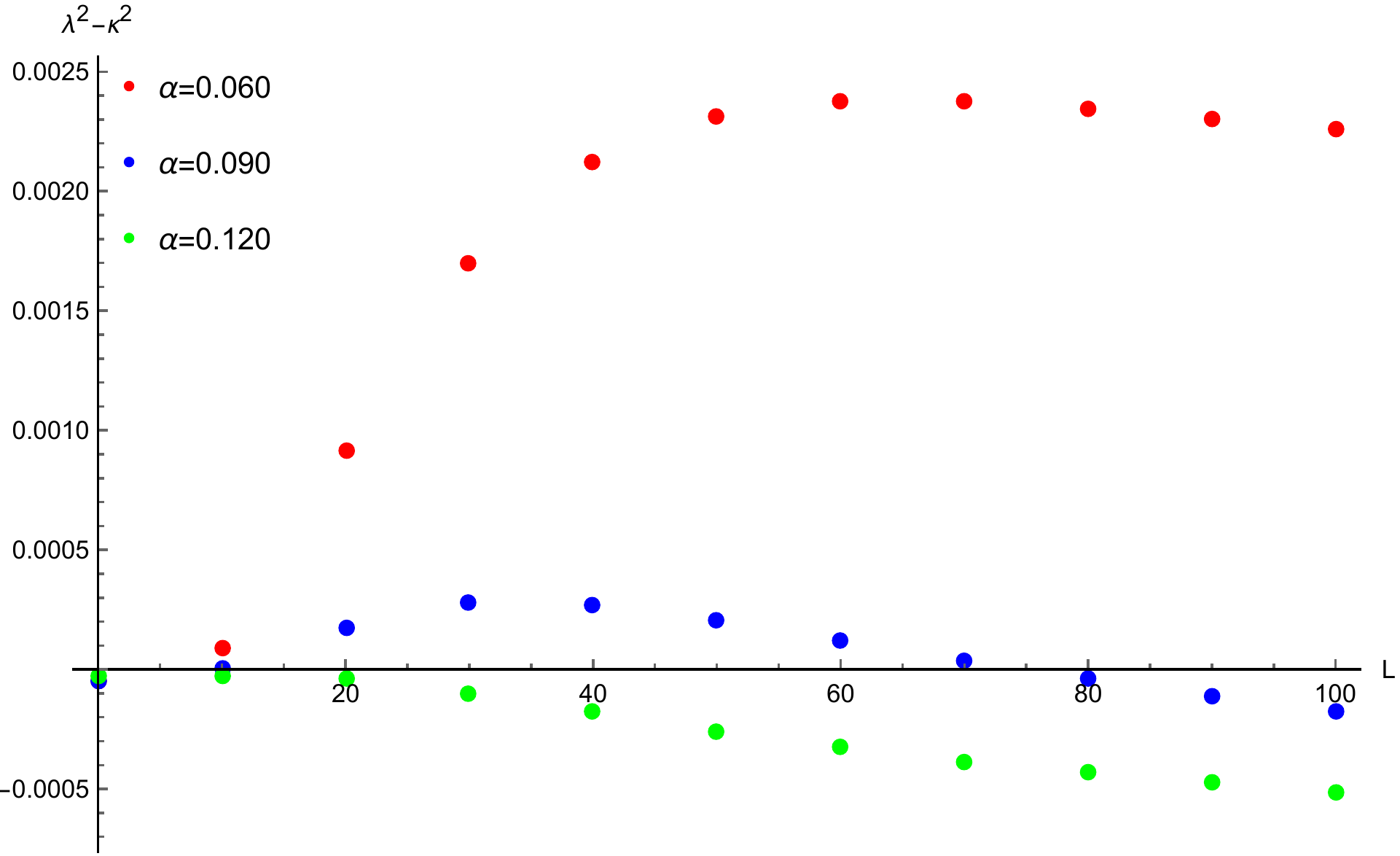}
\caption{The influence of the angular momentum on the chaos bound, where $\omega=-\frac{2}{3}$ and $Q=0.99$. The violation of the bound occurs at $L>7.35$ when $\alpha=0.060$ and at $10.00<L<75.10$ when $\alpha=0.090$. There is no violation when $\alpha=0.120$.}
\end{figure}

We insert the expressions of $F(r)$, $N(r)$ and $D(r)$ into Eq. (\ref{eq2.10}), and then numerically calculate the values of the Lyapunov exponent at the equilibrium orbits $r=r_0$ and the surface gravity. The chaos bound is analyzed in Figure 1-Figure 4. From the figures, the following phenomena are found.

1. When the values of the state parameter, normalization factor and angular momentum are fixed, different values of the electric charge produce different values of the Lyapunov exponent. When the electric charge is relatively small, no matter how the angular momentum is taken, the bound can not be violated. The violation of the bound only occurs when the electric charge is large enough.

2. The angular momentum of the particle affects the bound. When the value of the state parameter is fixed, and the electric charge is large enough, the angular momentum of the some specific values violates the bound. For example, the violation occurs when $\alpha= 0.020$, $L > 7.25$ and $\alpha= 0.060$, $9.27<L<52.41$ in Figure 1. A maximum value of $\lambda^2-\kappa^2$ occurs at $L =26.00$ when $\alpha= 0.060$.

3. The normalization factor affects the bound. When the values of the state parameter, electric charge and angular momentum are fixed, the maximum value of $\lambda^2-\kappa^2$ decreases with the increase of the normalization factor. It is more likely to cause the violation of the bound when the value of the normalization factor is relatively small. The value of the angular momentum corresponding to the maximum value of $\lambda^2-\kappa^2$ varies with the value of the normalization factor. In Figure 1 and Figure 3, we find that when $Q= 0.95$ and $\alpha > 0.096$, no matter how the value of the angular momentum increases, there is no violation. This phenomenon also occurs in Figure 2 and Figure 4.

4. The state parameter affects the bound. When $Q = 0.99$ and $\alpha = 0.12$, the angular momentum taking the specific values ($10.51<L<58.95$) leads to that the violation occurs at $\omega = - \frac{1}{2}$ and does not appears at $\omega = - \frac{2}{3}$.

5. The values of $\lambda^2-\kappa^2$ described by the red points in the figures are greater than $0$. In fact, there is $\lambda^2-\kappa^2 <0$ when the value of the angular momentum increases to a certain value, which shows that the bound is not violated in these regions.

Therefore, the state parameter, normalization factor and electric charge and the angular momentum jointly affect the chaos bound. The angular momentum of the particle plays an important role in the violation of the bound. When the value of the angular momentum is small, the equilibrium orbit is very close to the event horizon. From the figures, it is found that the bound is violated at a certain distance from the horizon and not violated in the near-horizon region. In \cite{ZLL}, the authors found the violation of the bound in the near-horizon regions through the Taylor expansions of the metric functions on the horizon. A sufficiently large charge of the black hole also causes the violation in the near-horizon region \cite{LG2}. When the charge mass ratio of the particle is large enough, we may get a result consistent with them.

\section{Conclusions}

In this paper, we investigated the chaos bound in the near-horizon region and at a certain distance from the horizon of the charged Kiselev black hole. The angular momentum of the charged particle plays an important role in the investigation. It affects not only the value of the Lyapunov exponent, but also the position of the equilibrium orbit. By fixing the charge mass ratio and changing the value of the angular momentum, we got the position which can be in the near-horizon region and at a certain distance from the horizon. Our result shows that when the electric charge is large enough and the angular momentum takes the specific values, the bound is violated at a certain distance from the horizon, and there is no violation in the near-horizon region. The small value of the normalization factor is more likely to cause the violation.


\begin{thebibliography}{99}                                                                                               %

\bibitem{HS1}
S.H. Shenker and D. Stanford, \emph{Black holes and the butterfly effect}, \emph{JHEP} \textbf{1403} (2014) 067.

\bibitem{HS2}
S.H. Shenker and D. Stanford, \emph{Stringy effects in scrambling}, \emph{JHEP} \textbf{1405} (2015) 132.

\bibitem{RRP}
R.R. Poojary, \emph{BTZ dynamics and chaos}, \emph{JHEP} \textbf{2003} (2020) 048.

\bibitem{JKY}
V. Jahnkea, K.Y. Kim and J. Yoon, \emph{On the chaos bound in rotating black holes}, \emph{JHEP} \textbf{1905} (2019) 037.

\bibitem{LR}
Y. Liu, A. Raju, \emph{Quantum chaos in topologically massive gravity}, \emph{JHEP} \textbf{2012} (2020) 027.

\bibitem{MSS}
J. Maldacena, S.H. Shenker and D. Stanford, \emph{A bound on chaos}, \emph{JHEP} \textbf{1608} (2016) 106.

\bibitem{PR}
J. Polchinski and V. Rosenhaus, \emph{The spectrum in the Sachdev-Ye-Kitaev model}, \emph{JHEP} \textbf{1604} (2016) 001.

\bibitem{MS}
J. Maldacena and D. Stanford, \emph{Remarks on the Sachdev-Ye-Kitaev model}, \emph{Phys. Rev.} \textbf{D 94} (2016) 106002.

\bibitem{ADNT}
M. Alishahiha, A. Davody, A. Naseh, S.F. Taghavi, \emph{On butterfly effect in higher derivative gravities}, \emph{JHEP} \textbf{1611} (2016) 032.

\bibitem{GHS}
G. Gur-Ari, M. Hanada and S.H. Shenker, \emph{Chaos in classical D0-brane mechanics}, \emph{JHEP} \textbf{1602} (2016) 091.

\bibitem{GZZ}
D. Giataganas, L.A. Pando Zayas and K. Zoubos, \emph{On marginal deformations and non-integrability}, \emph{JHEP} \textbf{1401} (2014) 129.

\bibitem{AKY}
Y. Asano, D. Kawai and K. Yoshida \emph{Chaos in the BMN matrix model}, \emph{JHEP} \textbf{1606} (2016) 191.

\bibitem{JASA}
A. Jawad, F. Ali, M.U. Shahzad and G. Abbas, \emph{Dynamics of particles around time conformal Schwarzschild black hole}, \emph{Eur. Phys. J.} \textbf{C 76} (2016) 586.

\bibitem{GZ}
D. Giataganas and K. Zoubos, \emph{Non-integrability and chaos with unquenched flavor}, \emph{JHEP} \textbf{1710} (2017) 042.

\bibitem{CFCZZ}
X. Chen, R.H. Fan, Y.M. Chen, H. Zhai and P.F. Zhang, \emph{Competition between chaotic and nonchaotic phases in a quadratically coupled Sachdev-Ye-Kitaev Model}, \emph{Phys. Rev. Lett.} \textbf{119} (2017) 207603.

\bibitem{HT}
K. Hashimoto and N. Tanahashi, \emph{Universality in chaos of particle motion near black hole horizon}, \emph{Phys. Rev.} \textbf{D 95} (2017) 024007.

\bibitem{KS}
A. Kitaev and S.J. Suh, \emph{The soft mode in the Sachdev-Ye-Kitaev model and its gravity dual}, \emph{JHEP} \textbf{1805} (2018) 183.

\bibitem{BLPV}
J.D. Boer, E. Llabres, J.F. Pedraza and D. Vegh, \emph{Chaotic strings in AdS/CFT}, \emph{Phys. Rev. Lett.} \textbf{120} (2018) 201604.

\bibitem{HG}
Y.C. Huang and Y.F. Gu, \emph{Eigenstate entanglement in the Sachdev-Ye-Kitaev model}, \emph{Phys. Rev.} \textbf{D} 100 (2019) 041901(R).

\bibitem{BINT}
M. Berkooz, M. Isachenkov, V. Narovlansky and G. Torrents, \emph{Towards a full solution of the large N double-scaled SYK model}, \emph{JHEP} \textbf{1903} (2019) 079.

\bibitem{LTW1}
X.B. Guo, K.K. Liang, B.R. Mu, P. Wang and H.T. Yang, \emph{Minimal length effects on motion of a particle in Rindler space}, \emph{Chin. Phys.} \textbf{C} 45 (2021) 023115.

\bibitem{CMBWZ}
V. Cardoso, A.S. Miranda, E. Berti, H. Witek and V.T. Zanchin, \emph{Geodesic stability, Lyapunov exponents and quasinormal modes}, \emph{Phys. Rev.} \textbf{D 79} (2009) 064016.

\bibitem{MWZ}
D.Z. Ma, J.P. Wu and J.F. Zhang, \emph{Chaos from the ring string in a Gauss-Bonnet black hole in $AdS_5$ space}, \emph{Phys. Rev.} \textbf{D 89} (2014) 086011.

\bibitem{PP1}
P. Pradhan, \emph{Stability analysis and quasinormal modes of Reissner-Nordstrom space-time via Lyapunov exponent}, \emph{Pramana} \textbf{87} (2016) 5.

\bibitem{PP2}
P. Pradhan, \emph{Circular geodesics in tidal charged black hole}, \emph{Int. J. Geom. Meth. Mod. Phys.} \textbf{15} (2017) 1850011.

\bibitem{CWJ1}
S.B. Chen, M.Z. Wang and J.L. Jing, \emph{Chaotic motion of particles in the accelerating and rotating black holes spacetime}, \emph{JHEP} \textbf{1609} (2016) 082.

\bibitem{WCJ2}
M.Z. Wang, S.B. Chen and J.L. Jing, \emph{Chaos in the motion of a test scalar particle coupling to the Einstein tensor in Schwarzschild-Melvin black hole spacetime}, \emph{Eur. Phys. J.} \textbf{C 77} (2017) 208.

\bibitem{MWZ1}
Y. Ling, P. Liu and J.P. Wu, \emph{Note on the butterfly effect in holographic superconductor models}, \emph{Phys. Lett.} \textbf{B 768} (2017) 288.

\bibitem{LTW2}
F.H. Lu, J. Tao and P. Wang, \emph{Minimal length effects on chaotic motion of particles around black hole horizon}, \emph{JCAP} \textbf{12} (2018) 036.

\bibitem{LW}
D. Li and X. Wu, \emph{Chaotic motion of neutral and charged particles in a magnetized Ernst-Schwarzschild spacetime}, \emph{Eur. Phys. J. Plus} \textbf{134} (2019) 96.

\bibitem{CKR}
G.J. Turiaci, \emph{An inelastic bound on chaos}, \emph{JHEP} \textbf{1907} (2019) 099.

\bibitem{DMM1}
S. Dalui, B.R. Majhi and P. Mishra, \emph{Presence of horizon makes particle motion chaotic}, \emph{Phys. Lett.} \textbf{B 788} (2019) 486 .

\bibitem{DMM2}
S. Dalui, B.R. Majhi and P. Mishra, \emph{Induction of chaotic fluctuations in particle dynamics in a uniformly accelerated frame}, \emph{Int. J. Mod. Phys.} \textbf{A 35} (2020) 2050081.

\bibitem{CKR1}
B. Craps, M. De Clerck, D. Janssens, V. Luyten and C. Rabideau, \emph{Lyapunov growth in quantum spin chains}, \emph{Phys. Rev.} \textbf{B 101} (2020) 174313.

\bibitem{CKR2}
B. Craps, S. Khetrapal and C. Rabideau, \emph{Chaos in CFT dual to rotating BTZ}, \emph{JHEP} \textbf{2111} (2021) 105.

\bibitem{ACO}
A. Addazi, S. Capozziello and S.D. Odintsov, \emph{Chaotic solutions and back hole shadow in $f(R)$ gravity}, \emph{Phys. Lett.} \textbf{B 816} (2021) 136257.

\bibitem{MPD}
M.H. Muoz-Arias, P.M. Poggi and I.H. Deutsch, \emph{Nonlinear dynamics and quantum chaos of a family of kicked p-spin models}, \emph{Phys. Rev. } \textbf{E 103} (2021) 052212.

\bibitem{HY}
X.L. Han and Z.D. Yu, \emph{Quantum chaos of the Bose-Fermi Kondo model at the intermediate temperature}, \emph{Phys. Rev.} \textbf{B 104} (2021) 085139.

\bibitem{GR}
D.J. Gross and V. Rosehips, \emph{Chaotic scattering of highly excited strings}, \emph{JHEP} \textbf{2105} (2021) 048.

\bibitem{CCKM1}
D. Chandorkar, S.D. Chowdhury, S, Kundu and S. Minwalla, \emph{Bounds on Regge growth of flat space scattering from bounds on chaos}, \emph{JHEP} \textbf{2105} (2021) 143

\bibitem{BL}
M. Blake and H. Liu, \emph{On systems of maximal quantum chaos}, \emph{JHEP} \textbf{2105} (2021) 229.

\bibitem{MRS}
M. Mondal, F. Rahaman and K.N. Singh, \emph{Lyapunov exponent ISCO and Kolmogorov Senai entropy for Kerr Kiselev black hole}, \emph{Eur. Phys. J.} \textbf{C 81} (2021) 84.

\bibitem{QCL}
L.C. Qu, J. Chen and Y.X. Liu, \emph{Chaos and complexity for inverted Harmonic oscillators}, [arXiv:2111.07351[hep-th]].

\bibitem{DG}
D. Giataganas, \emph{Chaotic motion near black hole and cosmological horizons}, [arXiv:2112.02081[hep-th]].

\bibitem{ZLL}
Q.Q. Zhao, Y.Z. Li and H. Lu, \emph{Static equilibria of charged particles around charged black holes: Chaos bound and its violations}, \emph{Phys. Rev.} \textbf{D 98} (2018) 124001.

\bibitem{KG1}
N. Kan and B. Gwak, \emph{Bound of Lyapunov exponent in Kerr-Newman black holes via charged particle}, \emph{Phys. Rev.} \textbf{D 105} 026006 (2022).

\bibitem{KG2}
B. Gwak, N. Kan, B.H. Lee and H. Lee, \emph{Violation of bound on chaos for charged probe in Kerr-Newman-AdS black hole}, [arXiv:2203.07298[gr-qc]].

\bibitem{LG1}
Y.Q. Lei, X.H. Ge and C. Ran, \emph{Chaos of particle motion near a black hole with quasitopological electromagnetism}, \emph{Phys. Rev.} \textbf{D 104} (2021) 046020.

\bibitem{LG2}
Y.Q. Lei and X.H. Ge, \emph{Circular motion of charged particles near charged black hole}, [arXiv:2111.06089[hep-th]].

\bibitem{BL1}
V.V. Kiselev, \emph{Quintessence and black holes}, \emph{Class. Quant. Grav.} \textbf{20} (2003) 1187.

\bibitem{MV1}
M. Visser, \emph{The Kiselev black hole is neither perfect fluid, nor is it quintessence}, \emph{Class. Quant. Grav.} \textbf{37} (2020) 045001.

\bibitem{MV2}
P. Boonserm, T. Ngampitipan, A. Simpson and M. Visser, \emph{Decomposition of total stress-energy for the generalised Kiselev black hole}, \emph{Phys. Rev.} \textbf{D 101} (2020) 024022.

\bibitem{BL3}
H. Shah, Z. Ahmad and H.H. Shah, \emph{Quintessence background for 4D Einstein-Gauss-Bonnet black holes}, \emph{Phys. Lett.} \textbf{B 818} (2021) 136383.

\bibitem{BL4}
M.S. Ali, F. Ahmed and S.G. Ghosh, \emph{Black string surrounded by a static anisotropic quintessence fluid}, \emph{Annals Phys.} \textbf{412} (2020) 168024.

\bibitem{TD2}
H.C. Kim, B.H. Lee, W. Lee and Y. Lee, \emph{Rotating black holes with an anisotropic matter field}, \emph{Phys. Rev.} \textbf{D 101} (2020) 064067.

\bibitem{OFH}
D. Ovchinnikov, M.U. Farooq, I. Hussain, A. Abdujabbarov, B. Ahmedov and Z. Stuchlík, \emph{Epicyclic oscillations of test particles near marginally stable circular orbits around charged Kiselev black holes}, \emph{Phys. Rev.} \textbf{D 104} (2021) 063027.

\bibitem{YJBH}
A. Younas, M. Jamil, S. Bahamonde and S. Hussain, \emph{Strong gravitational Lensing by Kiselev black hole}, \emph{Phys. Rev.} \textbf{D 92} (2015) 084042.

\bibitem{CFO}
B. Cuadros-Melgar, R.D.B. Fontana and J. de Oliveira, \emph{Superradiance and instabilities in black holes surrounded by anisotropic fluids}, \emph{Phys. Rev.} \textbf{D 104} (2021) 104039.

\bibitem{KO1}
I. Koga and Y. Ookouchi, \emph{Catalytic creation of a bubble universe induced by quintessence in five dimensions}, \emph{Phys. Rev.} \textbf{D 104} (2021) 126015.

\bibitem{KO2}
M. Rizwan, M. Jamil and A.Z. Wang, \emph{Distinguishing rotating Kiselev black hole from naked singularity using spin precession of test gyroscope}, \emph{Phys. Rev.} \textbf{D 98} (2018) 024015. Erratum: \emph{Phys. Rev.} \textbf{D 100} (2019) 029902.

\bibitem{BL6}
S.B. Chen, B. Wang and R.K. Su, \emph{Hawking radiation in a d-dimensional static spherically symmetric black hole surrounded by quintessence}, \emph{Phys. Rev.} \textbf{D 77} (2008) 124011.

\bibitem{GM1}
M. Azreg-Ainou and M.E. Rodrigues, \emph{Thermodynamical, geometrical and Poincare methods for charged black holes in presence of quintessence}, \emph{JHEP} \textbf{1309} (2013) 146.

\bibitem{GM2}
K. Ghaderi and B. Malakolkalami, \emph{Thermodynamics of the Schwarzschild and the Reissner-Nordstr$\ddot{o}$m black holes with quintessence}, \emph{Nucl. Phys. } \textbf{B 903} (2016) 10.









\end{thebibliography}
\end{document}